\documentclass{ws-ijmpa}
\usepackage[super,compress]{cite}
\usepackage{graphicx}
\usepackage{hyperref}
\usepackage{bbm}
\usepackage{amsmath}
\usepackage{amssymb}
\usepackage{doi}
\newcommand{\req}[1]{Eq.\,(\ref{#1})}
\newcommand{\rsec}[1]{Sec.\,{\ref{#1}}}

\newcommand{\orcidicon}{\includegraphics[width=0.32cm]{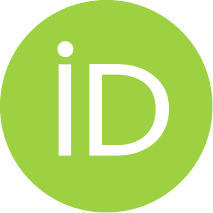}}
\newcommand{\orc}[1]{\href{https://orcid.org/#1}{\orcidicon}}

\newcommand{\orcA}{0000-0001-8217-1484}
\newcommand{\orcB}{0000-0001-5038-8427}
\newcommand{\orcC}{0000-0001-5474-2649}

\begin{document}
\markboth{J. Rafelski, A. Steinmetz, and C. T. Yang}{Dynamic fermion flavor mixing through transition dipole moments}

\catchline{}{}{}{}{}
%
%%%%%%%%%%%%%%%%%%%%%%%%%%%%%%%%%%%%%%%%%%%%%%%%%%%%%%%%%%%%%%%%%%%%

\title{Dynamic fermion flavor mixing through transition dipole moments}

\author{Johann Rafelski\orc{\orcA}, Andrew Steinmetz\orc{\orcC}, and Cheng Tao Yang\orc{\orcB}}

\address{Department of Physics, The University of Arizona, Tucson, AZ 85721, USA}
\maketitle

\begin{history}
\received{27 September 2023}
%\revised{Day Month Year}
\end{history}

\begin{abstract}
We show that Majorana neutrino flavor mixing can be driven by transition dipole moments in the presence of external electromagnetic fields. We demonstrate the sensitivity of the rotation mixing matrix to strong fields obtaining dynamical mass eigenstates in the two-flavor model. The three-flavor case, and extensions to the quark sector, are introduced.

\end{abstract}

\ccode{PACS numbers: 14.60.Lm, 14.60.Pq, 13.15.+g, 13.40.Em}

%%%%%%%%%%%%%%%%%%%%%%%%%%%%%%%%%%%%
\section{Introduction}
\label{sec:intro}
%%%%%%%%%%%%%%%%%%%%%%%%%%%%%%%%%%%%

Advancement of the understanding of neutrino physical properties attracts much interest today. Neutrinos are very abundant in the Universe; they were a dominant form of energy density in the Universe for much of its history, and they influence stellar and supernova evolution. As neutrinos are naturally massless in the standard model, the observed flavor oscillation signal non-vanishing neutrino mass. This suggests that neutrinos could provide a window to explore Beyond Standard Model (BSM) physics. This also motivates intense efforts to determine whether they are Dirac-type or Majorana-type fermions, with the latter serving as their own antiparticle.

We study the connection between Majorana neutrino transition magnetic dipole moments~\cite{Fujikawa:1980yx,Shrock:1980vy,Shrock:1982sc} and neutrino flavor oscillation. Neutrino electromagnetic (EM) properties have been considered before~\cite{Schechter:1981hw,Giunti:2014ixa,Popov:2019nkr,Chukhnova:2019oum}, including the effect of oscillation~\cite{Lim:1987tk,Akhmedov:1988uk,Pal:1991pm,Elizalde:2004mw,Akhmedov:2022txm} in magnetic fields. The case of transition moments has the mathematical characteristics of an off-diagonal mass, which is distinct from normal direct dipole moment behavior. EM field effects are also distinct from weak interaction remixing within matter, {\it i.e.\/} the Mikheyev-Smirnov-Wolfenstein effect~\cite{Wolfenstein:1977ue,Mikheyev:1985zog,Smirnov:2003da}.

For the case of two nearly degenerate neutrinos with transition moments, we determine in an explicit manner their eigenstates in the presence of EM fields. We obtain an EM-mass basis, distinct from flavor and free-particle mass basis, which mixes flavors as a function of EM fields. Moreover, we show solutions relating to full EM field tensor, which result in covariant expressions allowing for both magnetic and electrical fields we have not seen considered before. As neutrinos are electrically neutral, they have no intrinsic magnetic moment due to their spin; therefore any nonzero dipole moment is anomalous.

An anomalous magnetic moment (AMM) can be introduced into the neutrino Lagrangian via a Pauli term~\cite{Steinmetz:2018ryf,Itzykson:1980rh,Schwartz:2014sze}. Noteworthy for Majorana neutrinos, they can only possess transition magnetic moments which couple different flavors electromagnetically and do not violate CPT symmetry. Transition moments break lepton number conservation, therefore neutrino flavors could be remixed when exposed to strong EM fields.

The size of the neutrino magnetic dipole moment can be constrained as follows: The lower bound is found by higher order standard model interactions with the minimal extension of neutrino mass $m_{\nu}$ included~\cite{Fujikawa:1980yx,Shrock:1980vy,Shrock:1982sc}. The upper bound is derived from reactor, solar, and astrophysical experimental observations~\cite{Giunti:2015gga,Canas:2015yoa,Studenikin:2016ykv,AristizabalSierra:2021fuc}. The bounds are expressed in terms of the electron Bohr magneton $\mu_{B}$ as
\begin{align}
\label{bound:1}
\frac{e\hbar G_{F}m_{\nu}c^{2}}{8\pi^{2}\sqrt{2}} \sim 10^{-20}\mu_{B}<\mu_{\nu}^\mathrm{eff}<10^{-10}\mu_{B}\,,\qquad\mu_{B}=\frac{e\hbar}{2m_{e}}\;,
\end{align}
where $G_{F}$ is the Fermi constant and $\mu_{\nu}^\mathrm{eff}$ is the characteristic size of the neutrino magnetic moment. In~\req{bound:1}, the lower bound was estimated using a characteristic mass of $m_{\nu}\sim0.1~\mathrm{eV}$. From cosmological studies, the sum of neutrino masses is estimated~\cite{Planck:2018vyg} to be $\sum_{i}m_{i}<0.12$~eV; the electron (anti)neutrino mass is bounded~\cite{KATRIN:2021uub} by $m_{e}^{\nu}<0.8$~eV.

We discuss  the standard flavor mixing method and explore the Lagrangian density containing both Majorana mass and transition dipole moments in~\rsec{sec:nuflavor}.  In~\rsec{sec:numoment} we discuss the properties of the relativistic Pauli dipole. The two-flavor neutrino model is evaluated explicitly in~\rsec{sec:nutoy} where the remixed electromagnetic-mass eigenstates are obtained. The strong field (degenerate mass) and weak field limits are explored. While we focus our analysis on Majorana neutrinos, we note that the techniques in this work can be extended to apply to Dirac fermions in general; therefore they may be of interest also in the quark and charged lepton sector. Our results and future research outlook are described in~\rsec{sec:conclusions}.

%%%%%%%%%%%%%%%%%%%%%%%%%%%%%%%%%%%%%%%
\section{Neutrino flavor mixing and electromagnetic fields}
\label{sec:nuflavor}
%%%%%%%%%%%%%%%%%%%%%%%%%%%%%
%%%%%%%%%%%%%%%%%%%%%%%%%%%%%%%%%%%%%%%
\subsection{Standard flavor mixing method}
\label{sec:numethod}
%%%%%%%%%%%%%%%%%%%%%%%%%%%%%
Oscillation of neutrino flavors observed in experiment is in general interpreted as being due to a difference in neutrino mass and flavor eigenstates. This misalignment between the two representations is described as rotation of the neutrino flavor $N$-vector where $N=3$ is the observed number of generations. The unitary mixing matrix $V_{\ell k}$ allows for the change of basis between mass $(k)$ and flavor $(\ell)$ eigenstates via the transform 
\begin{alignat}{1}
\label{basis:1} \nu_{\ell}=\sum_{k=1}^{3}V_{\ell k}\nu_{k}\,\rightarrow
\begin{pmatrix}
\nu_{e}\\
\nu_{\mu}\\
\nu_{\tau}
\end{pmatrix}=
\begin{pmatrix}
V_{e1} & V_{e2} & V_{e3}\\
V_{\mu1} & V_{\mu2} & V_{\mu3}\\
V_{\tau1} & V_{\tau2} & V_{\tau3}
\end{pmatrix}
\begin{pmatrix}
\nu_{1}\\
\nu_{2}\\
\nu_{3}
\end{pmatrix}\,,
\end{alignat}
where $\nu_{\ell}$ is the neutrino state four-spinor written in the flavor basis, while in the mass basis we use $\nu_{k}$ with $k\in1,2,3$. Hereafter we will use implied summation over repeated flavor indices. Spinor indices will be suppressed.

The parameterization of the components of the mixing matrix depends on the Dirac or Majorana-nature of the neutrinos. First we recall the Dirac neutrino mixing matrix $U_{\ell k}$ in the standard parameterization~\cite{Schwartz:2014sze} 
\begin{alignat}{1}
\label{rotation:1} U_{\ell k} =
\begin{pmatrix}
 c_{12}c_{13} & s_{12}c_{13} & s_{13}e^{-i\delta}\\
 -s_{12}c_{23} - c_{12}s_{13}s_{23}e^{i\delta} & c_{12}c_{23} - s_{12}s_{13}s_{23}e^{i\delta} & c_{13}s_{23}\\
 s_{12}s_{23} - c_{12}s_{13}c_{23}e^{i\delta}& -c_{12}s_{23} - s_{12}s_{13}c_{23}e^{i\delta} & c_{13}c_{23}
\end{pmatrix}\,,
\end{alignat}
where $c_{ij} = \mathrm{cos}(\theta_{ij})$ and $s_{ij} = \mathrm{sin}(\theta_{ij})$. In this convention, the three mixing angles $(\theta_{12}, \theta_{13}, \theta_{23})$ are understood to be the Euler angles for generalized rotations and $\delta$ is the CP-violating complex phase. 

For the Majorana case we must allow a greater number of complex phases: Majorana neutrinos allow up to two additional complex phases $\rho$ and $\sigma$, which along with $\delta$, participate in CP-violation. A parameterization is achieved by introducing an additional phase matrix $P_{kk'}$
\begin{alignat}{1}
\label{phases:1} &V_{\ell k} = U_{\ell k'}P_{k'k}\,,\\
\label{phases:3} &P_{kk'} = \mathrm{diag}(e^{i\rho},e^{i\sigma},1)\,.
\end{alignat}
The mixing matrix $V_{\ell k}$ defined in~\req{phases:1} can then be used to transform the symmetric mass matrix $M_{\ell\ell'}$ from the flavor basis into the diagonal mass basis 
\begin{align}
\label{diag:1}
V_{\ell k}^{T}M_{\ell\ell'}V_{\ell'k'} = M_{kk'} = m_{k}\delta_{kk'} = \mathrm{diag}(m_{1},m_{2},m_{3})\,.
\end{align}
We note that there are many interesting models for mass matrices which were pioneered by Fritzsch and Xing~\cite{Fritzsch:1995dj,Fritzsch:1998xs,Fritzsch:1999ee,Xing:2000ik} in the leptonic sector. The masses $m_{k}$ are taken to be real and positive labelling the free propagating states of the three neutrinos.

%%%%%%%%%%%%%%%%%%%%%%%%%%%%%%%%%%%%%%%
\subsection{Majorana neutrino with dipole transition moment}
\label{sec:nuaction}
%%%%%%%%%%%%%%%%%%%%%%%%%%%%%
Turning now to the action, the Majorana mass term in the Lagrangian can be written in the flavor basis as
\begin{alignat}{1}
\label{mass:1} -\mathcal{L}_{\mathrm{mass}}^{\mathrm{Maj.}}=\frac{1}{2}\bar\nu_{\ell}M_{\ell\ell'}\nu_{\ell'}\,,\qquad
M_{\ell\ell'}^{T}=M_{\ell\ell'}\,,
\end{alignat}
where the Majorana fields are written as $\nu=\nu_{L}+C(\bar\nu_{L})^{T}$ and $\bar\nu=\nu^{\dag}\gamma_{0}$ is the Dirac adjoint. The field $\nu_{L}$ refers to left-handed chiral four-component spinors. The charge conjugation matrix $C$ is defined in the usual way in Ref.~\refcite{Itzykson:1980rh}, p.692. Charged conjugated fields are written as $\nu^{c}=C(\bar\nu)^{T}$. The Majorana mass matrix is symmetric, due to the anticommuting nature of the neutrino fields $\bar\nu\nu=-\nu^{T}\bar\nu^{T}$. It is in general complex~\cite{Adhikary:2013bma,giunti2007fundamentals}, though it will be taken to be fully real in this work. This is also why we have not shown the Hermitian conjugate in~\req{mass:1}.

Given these conventions, we can extend our consideration to include the electromagnetic interaction of neutrinos, which is possible if neutrinos are equipped with a magnetic moment matrix $\mu_{\ell\ell'}$. We allow for a fixed \emph{external} electromagnetic field tensor $F^{\alpha\beta}_\mathrm{ext}(x^{\mu})$, which imparts a force on the neutrino fields. We emphasize that $F^{\alpha\beta}_\mathrm{ext}$ is not dynamical in our formulation; it  consists of real functions over four-position and does not contain field operators.

We generalize the AMM Pauli spin-field Lagrangian to account for the Majorana fields in the flavor basis following the approach of Ref.~\refcite{Veltman:1997am}
\begin{align}
\label{moment:1}
-\mathcal{L}_{\mathrm{AMM}}^\mathrm{Maj.}=\frac{1}{2}\bar\nu_{\ell}\left(\mu_{\ell\ell'}\frac{1}{2}\sigma_{\alpha\beta}F^{\alpha\beta}_\mathrm{ext}\right)\nu_{\ell'}\,.
\end{align}
The operator $\sigma_{\alpha\beta}$ is the $4\times 4$ spin tensor defined by the commutator of the gamma matrices. We would like to point out some interesting features of the Pauli term most notably that the spin tensor itself is not Hermitian with
\begin{align}
\label{notherm:1}
\sigma_{\alpha\beta}^{\dag} = \gamma_{0}\sigma_{\alpha\beta}\gamma_{0}\,.
\end{align}
However, the conjugate of the Lagrangian term in~\req{moment:1},
\begin{align}
\left(\nu^{\dag}\gamma_{0}\sigma_{\alpha\beta}F^{\alpha\beta}_\mathrm{ext}\nu\right)^{\dag} = \nu^{\dag}\sigma_{\alpha\beta}^{\dag}F^{\alpha\beta}_\mathrm{ext}\gamma_{0}\nu = \nu^{\dag}\gamma_{0}\sigma_{\alpha\beta}F^{\alpha\beta}_\mathrm{ext}\nu\,,
\end{align}
is Hermitian. More about the spin tensor's properties will be elaborated on in~\rsec{sec:numoment}.

The Majorana magnetic moment matrix acts in flavor space. It satisfies the following constraints~\cite{Giunti:2014ixa} for CPT symmetry reasons and the anticommuting nature of fermions
\begin{alignat}{1}
\label{props:1}
\mu_{\ell\ell'}^{\dag}=\mu_{\ell\ell'}\,,\qquad
\mu_{\ell\ell'}^{T}=-\mu_{\ell\ell'}\,,
\end{alignat}
{\it i.e.\/} the AMM matrix $\mu_{\ell\ell'}$ is Hermitian and fully anti-symmetric. This requires that the transition magnetic moment elements are purely imaginary while all diagonal AMM matrix elements vanish
\begin{align}
\label{mu:1}
\mu_{\ell\ell'}=
\begin{pmatrix}
\mu_{ee} & \mu_{e\mu} & \mu_{e\tau} \\
\mu_{\mu e} & \mu_{\mu\mu} & \mu_{\mu\tau} \\
\mu_{\tau e} & \mu_{\tau\mu} & \mu_{\tau\tau}
\end{pmatrix}\xrightarrow{\mathrm{Majorana}}
\mu_{\ell\ell'}=
\begin{pmatrix}
0 & i\mu_{e\mu} & -i\mu_{e\tau} \\
-i\mu_{e\mu} & 0 & i\mu_{\mu\tau} \\
i\mu_{e\tau} & -i\mu_{\mu\tau} & 0
\end{pmatrix}\,.
\end{align}

We can combine the mass term in~\req{mass:1} and AMM contribution in~\req{moment:1} into a single effective Lagrangian
\begin{align}
\label{massmom:1}
\mathcal{L}_\mathrm{eff}^\mathrm{Maj.} &= \mathcal{L}_\mathrm{kinetic}^\mathrm{Maj.} + \mathcal{L}_\mathrm{mass}^\mathrm{Maj.} + \mathcal{L}_\mathrm{AMM}^\mathrm{Maj.}\,,
\end{align}
\begin{align}
\label{massmom:2}
\mathcal{L}_\mathrm{eff}^\mathrm{Maj.} &= \mathcal{L}_\mathrm{kinetic}^\mathrm{Maj.} - \frac{1}{2}\bar\nu_{\ell}\left(M_{\ell\ell'}+\mu_{\ell\ell'}\frac{1}{2}\sigma_{\alpha\beta}F^{\alpha\beta}_\mathrm{ext}\right)\nu_{\ell'}\;.
\end{align}
\req{massmom:2} is our working Lagrangian. We define the generalized mass-dipole matrix $\mathcal{M}_{\ell\ell'}$ present in~\req{massmom:2} as
\begin{align}
\label{massmom:3}
\mathcal{M}_{\ell\ell'}(E,B)\equiv M_{\ell\ell'}+\mu_{\ell\ell'}\frac{1}{2}\sigma_{\alpha\beta}F^{\alpha\beta}_\mathrm{ext}\,,\qquad \mathcal{M}_{\ell\ell'}^{\dag}=\gamma_{0}\mathcal{M}_{\ell\ell'}\gamma_{0}\,.
\end{align}
The presence the EM interaction generated by the transition dipole moment is now understood to create  an effective mass $m\rightarrow\widetilde m(E,B)$ inducing new EM dependence of the mixing matrix, leading to modifications seen in~\req{phases:1}. As neutrinos propagate as energy eigenstates, our objective is to recognize and understand  the effect of mass modification due to electromagnetic field; thus we consider the eigenvalues of~\req{massmom:3} rather than those seen in~\req{diag:1}.  The electromagnetic effect then contributes to the time-dependant oscillation among the free-particle mass eigenstates~\cite{Giunti:2014ixa}.

%%%%%%%%%%%%%%%%%%%%%%%%%%%%%%%%%%%%%%%
\subsection{Chiral properties of the relativistic Pauli dipole}
\label{sec:numoment}
%%%%%%%%%%%%%%%%%%%%%%%%%%%%%%%%%%%%%%%
The electromagnetic dipole behavior of the neutrino depends on the mathematical properties of the tensor product $\sigma_{\alpha\beta}F^{\alpha\beta}_\mathrm{ext}$. We prefer to work in the Weyl (chiral) spinor representation where the EM contribution is diagonal in spin space. Therefore, we evaluate the product $\sigma_{\alpha\beta}F^{\alpha\beta}_\mathrm{ext}$ in the Weyl representation following Feynman and Gell-mann\cite{Feynman:1958ty}, yielding
\begin{align}
\label{chiral:1}
-\frac{1}{2}\sigma_{\alpha\beta}F^{\alpha\beta}_\mathrm{ext}=
\begin{pmatrix}
\vec{\sigma}\cdot(\vec{B}+i\vec{E}/c) & 0\\
0 & \vec{\sigma}\cdot(\vec{B}-i\vec{E}/c)
\end{pmatrix}\equiv
\begin{pmatrix}
\vec{\sigma}\cdot\vec{f}_{+} & 0 \\
0 & \vec{\sigma}\cdot\vec{f}_{-}
\end{pmatrix}\,,
\end{align}
where we introduced the complex electromagnetic field form $\vec{f}_{\pm}=\vec{B}\pm i\vec{E}/c$ showing sensitivity to both magnetic and electric fields. As this expression is diagonal in the Weyl representation, it does not exchange handedness when acting upon a state. Since left and right-handed neutrinos are not remixed by transition magnetic moments, sterile right-handed neutrinos do not need to be introduced. We can also see explicitly in~\req{chiral:1} its non-Hermitian character, see~\req{massmom:2}, of the EM spin-field coupling. Specifically this is mirrored in the complex field's $\vec{f}_{\pm}$ relation to its complex conjugate $(\vec{f}_{\pm})^{*}=\vec{f}_{\mp}$. The complex EM fields have a Hermitian $(\vec{B})$ and anti-Hermitian $(i\vec{E})$ part.

For later convenience, we note how the EM invariants $\mathcal{S}$ and $\mathcal{P}$ help explain the AMM term 
\begin{align}
\label{invar:1}
\frac{1}{2}\left(\frac{1}{2}\sigma_{\alpha\beta}F^{\alpha\beta}_\mathrm{ext}\right)^{2}=
\begin{pmatrix}
\mathcal{S}+i\mathcal{P} & 0\\
0 & \mathcal{S}-i\mathcal{P}
\end{pmatrix}=\mathcal{S}-i\gamma_{5}\mathcal{P}\,,
\end{align}
\begin{align}
\mathcal{S}\equiv\frac{1}{2}\left(B^{2}-E^{2}/c^{2}\right)\,,\qquad
\mathcal{P}\equiv\vec{B}\cdot\vec{E}/c\,,\qquad
\frac{1}{2}\vec{f}_{\pm}\cdot\vec{f}_{\pm}=\mathcal{S}\pm i\mathcal{P}\,.
\end{align}
The combination of these invariants make up the eigenvalues of the Pauli term. Moreover, taking the product of $\vec{f}_{\pm}$ with its complex conjugate we find
\begin{align}
\label{cross:1}
\frac{1}{2}\left(\vec{\sigma}\cdot\vec{f}_{\pm}\right)\left(\vec{\sigma}\cdot\vec{f}_{\mp}\right)=T_\mathrm{ext}^{00}\mp \sigma_{i}T_\mathrm{ext}^{0i}\,,
\end{align}
where we recognize the stress-energy tensor $T_\mathrm{ext}^{\alpha\beta}$ component $T_\mathrm{ext}^{00}$ for field energy density, and $T_\mathrm{ext}^{0i}$ momentum density, respectively
\begin{align}
T_\mathrm{ext}^{00}=\frac{1}{2}\left(B^{2}+E^{2}/c^{2}\right)\,,\qquad
T_\mathrm{ext}^{0i}=\frac{1}{c}\varepsilon_{ijk}E_{j}B_{k}\,.
\end{align}
As we will see in~\rsec{sec:nutoy},~\req{cross:1} will appear in the EM-mass eigenvalues of our effective Lagrangian~\req{massmom:1}. Using the identity in~\req{chiral:1} and~\req{cross:1} we also find the interesting relationship
\begin{align}
\label{cross:2}
\frac{1}{2}\left(\frac{1}{2}\sigma_{\alpha\beta}F^{\alpha\beta}_\mathrm{ext}\right)\left(\frac{1}{2}\sigma_{\alpha\beta}F^{\alpha\beta}_\mathrm{ext}\right)^{\dag}=
\gamma_{0}\left(T_\mathrm{ext}^{00}\gamma_{0}+T_\mathrm{ext}^{0i}\gamma_{i}\right)\,.
\end{align}
Now that we have elaborated on the relevant EM field identities, we turn back to the magnetic dipole and flavor rotation problem.

%%%%%%%%%%%%%%%%%%%%%%%%%%%%%%%%%%%%%%%
\section{Two-flavor toy model of electromagnetic-mass mixing}
\label{sec:nutoy}
%%%%%%%%%%%%%%%%%%%%%%%%%%%%%%%%%%%%%%%
\subsection{Fundamentals of electromagnetic mixing method}
Considering experimental data on neutrino oscillations, it is understood that either the two lighter (normal hierarchy) or the two heavier (inverted hierarchy) neutrino states are close together in mass. If the electromagnetic properties of the neutrino do indeed lead to flavor mixing effects, then it is likely the closer pair of neutrino mass states that are most sensitive to the phenomenon we explore. In the spirit of Bethe~\cite{Bethe:1986ej}, we therefore explore the $N=2$ two generation $(\nu_{e},\nu_{\mu})$ toy model.

Following the properties established in~\req{props:1} and~\req{massmom:3} we write down the two generation mass and dipole matrices as
\begin{alignat}{1}
\label{mix:1} M_{\ell\ell'}= 
\begin{pmatrix}
m_{e}^{\nu} & {\delta m}\\
{\delta m} & m_{\mu}^{\nu}
\end{pmatrix}\,,\qquad
\mu_{\ell\ell'} = 
\begin{pmatrix}
0 & i\delta\mu\\
-i\delta\mu & 0
\end{pmatrix}\,.
\end{alignat}
The AMM coupling $\delta\mu$ is taken to be real with a pure imaginary coefficient. While the mass elements $(m_{e}^{\nu},m_{\mu}^{\nu},{\delta m})$ are generally complex, we choose in our toy model for them to be fully real
\begin{align}
\label{choice:1}
m_{e}^{\nu}=(m_{e}^{\nu})^{*}\,,\qquad
m_{\nu_{\mu}}=(m_{\mu}^{\nu})^{*}\,,\qquad
\delta m=\delta m^{*}\,,
\end{align}
making the mass matrix $M_{\ell\ell'}$ Hermitian. This allows us to more easily evaluate  the EM contributions to mixing, avoiding the complications arising from the mass matrix.

Using~\req{mix:1} and~\req{choice:1}, we write the mass-dipole matrix in~\req{massmom:3} in terms of $2\times2$ flavor components as
\begin{align}
\label{mix:2}
\mathcal{M}_{\ell\ell'} = 
\begin{pmatrix}
m_{e}^{\nu} & {\delta m}+i\delta\mu\sigma_{\alpha\beta}F^{\alpha\beta}_\mathrm{ext}/2\\
{\delta m}-i\delta\mu\sigma_{\alpha\beta}F^{\alpha\beta}_\mathrm{ext}/2 & m_{\mu}^{\nu}
\end{pmatrix}\,,\qquad
\mathcal{M}_{\ell\ell'}^{\dag}=\gamma_{0}\mathcal{M}_{\ell\ell'}\gamma_{0}\,.
\end{align}
As noted before, this matrix is not Hermitian due to the inclusion of the spin tensor. However, any arbitrary complex matrix can be diagonalized into its real eigenvalues $\lambda_{j}$ by the biunitary transform
\begin{align}
\label{biunitary:1}
W_{\ell j}^{\dag}\mathcal{M}_{\ell\ell'}Y_{\ell'j'}=\lambda_{j}\delta_{jj'}\,,
\end{align}
where $Y_{\ell j}$ and $W_{\ell j}$ are both unitary matrices. Taking the complex conjugate of~\req{biunitary:1}, we arrive at
\begin{align}
\label{biunitary:2}
(W_{\ell j}^{\dag}\mathcal{M}_{\ell\ell'}Y_{\ell'j'})^{\dag} = 
Y_{\ell j'}^{\dag}\gamma_{0}\mathcal{M}_{\ell\ell'}\gamma_{0}W_{\ell' j}=\lambda_{j}\delta_{jj'}\,,
\end{align}
\begin{align}
Y_{\ell j}=\gamma_{0}W_{\ell j}\rightarrow
W_{\ell j}^{\dag}\mathcal{M}_{\ell\ell'}\gamma_{0}W_{\ell'j'}=\lambda_{j}\delta_{jj'}\,. 
\end{align}
As $Y_{\ell j}$ and $W_{\ell j}$ are related by a factor of $\gamma_{0}$ based on the conjugation properties of~\req{mix:2}, this lets us eliminate $Y_{\ell j}$ and diagonalize using a single unitary matrix $W_{\ell j}$. The related matrix $\mathcal{M}_{\ell\ell'}\gamma_{0}$ is Hermitian
\begin{align}
\label{herm:1}
(\mathcal{M}_{\ell\ell'}\gamma_{0})^{\dag} = \mathcal{M}_{\ell\ell'}\gamma_{0}\,,
\end{align}
and also equivalent to the root of the Hermitian product of~\req{mix:2}
\begin{align}
(\mathcal{M}\mathcal{M}^{\dag})_{\ell\ell'} = \left((\mathcal{M}\gamma_{0})(\mathcal{M}\gamma_{0})\right)_{\ell\ell'}\,.
\end{align}
Therefore a suitable unitary transformation $W_{\ell j}$ rotates flavor $\ell$-states into magnetized mass $j$-states. The eigenvalues $\lambda_{j}^{2}$ of $(\mathcal{M}\mathcal{M}^{\dag})_{\ell\ell'}$ are the squares of both signs of the eigenvalues of $\mathcal{M}_{\ell\ell'}\gamma_{0}$. We write this property (with flavor indices suppressed) as
\begin{align}
W^{\dag}(\mathcal{M}\mathcal{M}^{\dag})W &= W^{\dag}(\mathcal{M}\gamma_{0})WW^{\dag}(\mathcal{M}\gamma_{0})W = \mathrm{diag}(\lambda_{1}^{2},\lambda_{2}^{2})\,.
\end{align}
We recognize $\lambda_{j}=\widetilde m_{j}(E,B)$ with $j\in1,2$ as the effective mass states which are EM-field dependant in this basis. 

%%%%%%%%%%%%%%%%%%%%%%%%%%%%%%%%%%%%%%%
\subsection{Separating electromagnetic-mass mixing}
\label{sec:zmixing}
%%%%%%%%%%%%%%%%%%%%%%%%%%%%%%%%%%%%%%%
The matrix $W_{\ell j}$ mixes flavor states into a new basis distinct from the free-particle case however this rotation must smoothly connect with the free-particle case in the limit that the electromagnetic fields go to zero. We proceed to evaluate $W_{\ell j}$, breaking the rotation into two separate unitary transformations: a) rotation $V_{\ell k}^{\dag}(\ell\rightarrow k)$ to free-particle mass; and b) rotation $Z_{kj}^{\mathrm{ext}\dag}(k\rightarrow j)$ to the EM-mass basis. Guided by~\req{basis:1} we write
\begin{align}
\label{zrot:1}
\nu_{j} = W^{\dag}_{\ell j}\nu_{\ell} = Z_{kj}^{\mathrm{ext}\dag}V_{\ell k}^{\dag}\nu_{\ell}\,.
\end{align}
In the limit that the EM fields go to zero, the electromagnetic rotation becomes unity $Z_{kj}^\mathrm{ext}\rightarrow\delta_{kj}$, thereby ensuring the EM-mass basis and the free-particle mass basis become equivalent. The rotation $Z_{kj}^\mathrm{ext}$ can then be interpreted as the external field forced rotation. While our argument above is done explicitly for the two generation case, it can be generalized to accommodate three generations of neutrinos as well.

According to~\req{diag:1}, the mass matrix in~\req{mix:1} can be diagonalized in the two generation case by a one parameter unitary mixing matrix $V_{\ell k}$ given by
\begin{align}
\label{rot:1}
V_{\ell k}(\theta)=
\begin{pmatrix}
\cos\theta & \sin\theta\\
-\sin\theta & \cos\theta
\end{pmatrix}\,.
\end{align}
For a real Hermitian $2\times 2$ mass matrix, the rotation matrix $V_{\ell k}$ is real and only depends on the rotation angle $\theta$. The explicit form of the EM-field related rotation $Z_{kj}^\mathrm{ext}$ introduced in~\req{zrot:1} is
\begin{align}
\label{zrot:2}
Z_{kj}^\mathrm{ext}(\omega,\phi)=
\begin{pmatrix}
\cos\omega & e^{i\phi}\sin\omega\\
-e^{-i\phi}\sin\omega & \cos\omega
\end{pmatrix}\,,\qquad
W_{\ell j}(\theta,\omega,\phi)=V_{\ell k}(\theta)Z_{kj}^\mathrm{ext}(\omega,\phi)\,,
\end{align}
where $Z_{kj}^\mathrm{ext}$ depends on the real angle $\omega$ and the complex phase $\phi$. The full rotation $W_{\ell j}$ therefore depends on three parameters when broken into the free-particle rotation and the EM rotation.

The eigenvalues of the original Hermitian mass matrix in~\req{mix:1} are given by
\begin{align}
\label{massroot:1}
m_{1,2}=\frac{1}{2}\left(m_{\nu_{e}}+m_{\nu_{\mu}}\mp\sqrt{|\Delta m_{0}|^{2}+4\delta m^{2}}\right)\,,\qquad
|\Delta m_{0}|=|m_{\nu_{\mu}}-m_{\nu_{e}}|\,.
\end{align}
We assign $m_{1}$ to the lower mass $(-)$ root and $m_{2}$ with the larger mass $(+)$ additive root. The rotation $\theta$ in~\req{rot:1} is then given by
\begin{align}
\label{massroot:2}
\sin2\theta=\sqrt{\frac{4\delta m^{2}}{|\Delta m_{0}|^{2}+4\delta m^{2}}}\,,\qquad
\cos2\theta=\sqrt{\frac{|\Delta m_{0}|^{2}}{|\Delta m_{0}|^{2}+4\delta m^{2}}}\,.
\end{align}

In our toy model, the off-diagonal imaginary transition magnetic moment  $\mu_{\ell\ell'}$ commutes with the real valued mixing matrix $V_{\ell k}$ and the following relations hold
\begin{align}
\label{commuting:1}
V_{\ell k}^{\dag}\mu_{\ell\ell'}V_{\ell' k'}=(V^{\dag}V)_{k\ell'}\mu_{\ell'k'}=\mu_{kk'}=
\begin{pmatrix}
0 & i\delta\mu\\
-i\delta\mu & 0
\end{pmatrix}\,.
\end{align}
We see that the Majorana transition dipoles in our model are off-diagonal in both flavor and mass basis. Therefore the real parameter unitary matrix in~\req{commuting:1} cannot rotate a pure imaginary matrix at least in the two flavor case. We apply the rotation in~\req{rot:1} to~\req{herm:1} yielding
\begin{align}
\label{herm:2}
V_{\ell k}^{\dag}(\mathcal{M}_{\ell\ell'}\gamma_{0})V_{\ell' k'} &= 
V_{\ell k}^{\dag}M_{\ell\ell'}\gamma_{0}V_{\ell' k'} +
V_{\ell k}^{\dag}(\mu_{\ell\ell'}\sigma_{\alpha\beta}\gamma_{0}F^{\alpha\beta}_\mathrm{ext}/2)V_{\ell' k'}\,,
\end{align}
\begin{align}
\label{herm:3}
V_{\ell k}^{\dag}(\mathcal{M}_{\ell\ell'}\gamma_{0})V_{\ell' k'} &= 
\begin{pmatrix}
m_{1}\gamma_{0} & i\delta\mu\sigma_{\alpha\beta}\gamma_{0}F^{\alpha\beta}_\mathrm{ext}/2\\
-i\delta\mu\sigma_{\alpha\beta}\gamma_{0}F^{\alpha\beta}_\mathrm{ext}/2 & m_{2}\gamma_{0}
\end{pmatrix}\equiv
\begin{pmatrix}
\mathcal{A} & i\mathcal{C}\\
-i\mathcal{C} & \mathcal{B}
\end{pmatrix}\,,
\end{align}
where we have defined implicitly the Hermitian elements $(\mathcal{A},\mathcal{B},\mathcal{C})$.  Applying now both rotations 
to~\req{herm:1} yields
\begin{align}
\label{herm:4}
W_{\ell j}^{\dag}(\mathcal{M}_{\ell\ell'}\gamma_{0})W_{\ell' j'} &= 
Z^{\mathrm{ext}\dag}\begin{pmatrix}
\mathcal{A} & i\mathcal{C}\\
-i\mathcal{C} & \mathcal{B}
\end{pmatrix}Z^\mathrm{ext}=\lambda_{j}\delta_{jj'}\,.
\end{align}
\req{herm:4} is therefore the working matrix equation which needs to be solved to identify the EM rotation parameters. As discussed before, this means that the rotation angle $\omega$ and the phase $\phi$ are in general functions of electromagnetic fields.

%%%%%%%%%%%%%%%%%%%%%%%%%%%%%%%%%%%%%%%
\subsection{Dynamic electromagnetic-mass eigenvalue results}
\label{sec:emmass}
%%%%%%%%%%%%%%%%%%%%%%%%%%%%%%%%%%%%%%%
We will now solve for the rotation parameters necessary to define the EM-mass basis which acts as a distinct propagating basis for neutrinos in external fields. Considering that the $j$-columns vectors $v_{k}^{(j)}$ of $Z_{kj}^\mathrm{ext}$ are eigenvectors for each $\lambda_{j}$
\begin{align}
\label{herm:5}
Z_{kj}^\mathrm{ext}=v_{k}^{(j)}=
\begin{pmatrix}
v^{1} & v^{2}
\end{pmatrix}\,,
\end{align}
\req{herm:4} has the meaning of an eigenvalue equation
\begin{align}
\label{herm:6}
\begin{pmatrix}
\mathcal{A} & i\mathcal{C}\\
-i\mathcal{C} & \mathcal{B}
\end{pmatrix}Z^\mathrm{ext}=
Z^\mathrm{ext}\begin{pmatrix}
\lambda_{1} & 0\\
0 & \lambda_{2}
\end{pmatrix}\rightarrow
\begin{pmatrix}
\mathcal{A} & i\mathcal{C}\\
-i\mathcal{C} & \mathcal{B}
\end{pmatrix}v^{(j)}=\lambda_{j}v^{(j)}\,.
\end{align}

Given the eigenvalue equation defined in~\req{herm:6}, we obtain the effective EM-masses as solutions to the characteristic polynomial
\begin{align}
\label{poly:1}
(\mathcal{A}-\lambda_{j}\gamma_{0})(\mathcal{B}-\lambda_{j}\gamma_{0})-\mathcal{C}^{2}=0\,,
\end{align}
which we obtained by taking the determinant of~\req{herm:6} over flavor but not spin space. It is useful to define the following identities for the off-diagonal element
\begin{align}
\label{poly:1a}
\mathcal{C}^{2} = 
\delta\mu^{2}\left(\frac{1}{2}\sigma_{\alpha\beta}F^{\alpha\beta}_\mathrm{ext}\right)\left(\frac{1}{2}\sigma_{\alpha\beta}F^{\alpha\beta}_\mathrm{ext}\right)^{\dag}=
2\delta\mu^{2}\gamma_{0}\left(T_\mathrm{ext}^{00}\gamma_{0}+T_\mathrm{ext}^{0i}\gamma_{i}\right)\,,
\end{align}
and for the diagonal elements
\begin{align}
(\mathcal{B}-\mathcal{A})^{2} = |m_{2}-m_{1}|^{2} = |\Delta m|^{2}\,,\qquad (\mathcal{A}+\mathcal{B})\gamma_{0} = m_{1} + m_{2}\,.
\end{align}
\req{poly:1a} was obtained using the expression in~\req{cross:2}. Because of the spinor behavior of each element, the eigenvalues are obtained with $\gamma_{0}$ coefficients.~\req{poly:1} therefore has the roots $\lambda_{1,2} = \widetilde m_{1,2}(E,B)$
\begin{align}
\label{poly:2}
\widetilde m_{1,2}(E,B)\! &=\! \frac{1}{2}\left(m_{1}\!+\!m_{2}\!\mp\!\sqrt{|\Delta m|^{2}\!+\!8\delta\mu^{2}\gamma_{0}\left(T_\mathrm{ext}^{00}\gamma_{0}+T_\mathrm{ext}^{0i}\gamma_{i}\right)}\right)\!,
\end{align}
\begin{align}
\label{poly:3}
\widetilde m_{1,2}(E,B)\! &=\! \frac{1}{2}\left(m_{1}\!+\!m_{2}\!\mp\!\sqrt{|\Delta m|^{2}\!+\!8\delta\mu^{2}\gamma_{0}\left(\gamma_{0}\frac{1}{2}\left(B^{2}\!+\!\frac{E^{2}}{c^{2}}\right)\!+\!\vec{\gamma}\!\cdot\!(\frac{\vec{E}}{c}\times\vec{B})\right)}\right)\!.
\end{align}
The EM-mass eigenstates $\widetilde m(E,B)$ depends on the energy density $T_\mathrm{ext}^{00}$ of the EM field and the spin projection along the EM momentum density $T_\mathrm{ext}^{0i}$. However the coefficient $\delta\mu^{2}$ is presumed to be very small, therefore the EM contribution only manifests in strong EM fields or where the free-particle case has very nearly or exactly degenerate masses, $\Delta m\to 0$. When the the electromagnetic fields go to zero, the EM-masses in~\req{poly:3} reduce as expected to the free-particle result.

The complex phase in~\req{zrot:2} has the value $\phi=\pi(n-1/2)$ with $n\in0,\pm1,\pm2...$ making the complex exponential in~\req{zrot:2} pure imaginary. Curiously, the phase is not field dependant, but tied to the fact that the Majorana moments are pure imaginary quantities. Complex phases in mixing matrices are generally associated with CP violation such as the Dirac phase $\delta$ in~\req{rotation:1} which suggests that CP violation in the neutrino sector can be induced in the presence of external EM fields. Further analysis of the three generation case is left to future work.

We note that the solution in~\req{poly:3} actually contain four distinct EM-mass eigenstates $\widetilde m_{j}^{s}(E,B)$ with the lower $(j=1)$ and the upper $(j=2)$ masses, and the additional spin splitting from the alignment $(s=+1)$, or anti-alignment $(s=-1)$, of the neutrino spin with the momentum density of the external EM field. Spin splitting vanishes for the pure electric or magnetic field cases. For good spin eigenstates $s\in\pm1$, we can rewrite~\req{poly:1a} with EM fields explicitly as
\begin{align}
\label{spinsplit:1}
\mathcal{C}^{2}_{s}(E,B)=2\delta\mu^{2}\left(\frac{1}{2}(B^{2}+E^{2}/c^{2})+s|\vec{E}/c\times\vec{B}|\right)\,.
\end{align}
The above expression within the square is positive definite; therefore~\req{spinsplit:1} is always real. Spin splitting requires that we consider separate rotations for each spin state as the rotation angle $\omega_{s}$ depends on the spin quantum number
\begin{align}
\label{zrot:3}
\sin2\omega_{s}=\sqrt{\frac{4\mathcal{C}_{s}^{2}}{|\Delta m|^{2}+4\mathcal{C}_{s}^{2}}}\,,\qquad
\cos2\omega_{s}=\sqrt{\frac{|\Delta m|^{2}}{|\Delta m|^{2}+4\mathcal{C}_{s}^{2}}}\,.
\end{align}
The expressions in~\req{zrot:3} are mathematically similar to that of the free-particle case written in~\req{massroot:2} in the two flavor generation model with the off-diagonal mass being replaced with the EM dependant quantity $\mathcal{C}_{s}$.

%%%%%%%%%%%%%%%%%%%%%%%%%%%%%%%%%%%%%%%
\subsection{Strong field (degenerate mass) and weak field limits}
\label{sec:nulimits}
%%%%%%%%%%%%%%%%%%%%%%%%%%%%%%%%%%%%%%%
The rotation angles in~\req{zrot:3} reveal two distinct limits where EM-masses are dominated by either: a) the intrinsic mass splitting $\mathcal{C}_{s}\ll|\Delta m|^{2}$ with $\omega_{s}\rightarrow0$ or b) the EM contribution $\mathcal{C}_{s}\gg|\Delta m|^{2}$ where $\omega_{s}\rightarrow\pi/4$.
For the first case where the masses are not degenerate or the fields are weak, we obtain the expansion
\begin{align}
\label{series:1}
\lim_{\mathcal{C}_{s}\ll|\Delta m|^{2}}\widetilde m_{1,2}^{s}(E,B)=\frac{1}{2}\left(m_{1}+m_{2}\mp|\Delta m|\left(1+\frac{2\mathcal{C}_{s}^{2}}{|\Delta m|^{2}}+\ldots\right)\right)\,,
\end{align}
which as stated before reduces to the free-particle case at lowest order.

In the opposite limit, where the masses are very nearly degenerate or fields are strong, the EM-mass eigenvalues in~\req{poly:3} can be approximated by the series
\begin{align}
\label{series:2}
\lim_{\mathcal{C}_{s}\gg|\Delta m|^{2}}\widetilde m_{1,2}^{s}(E,B)=\frac{1}{2}\left(m_{1}+m_{2}\mp2\mathcal{C}_{s}\left(1+\frac{|\Delta m|^{2}}{8\mathcal{C}_{s}^{2}}+\ldots\right)\right)
\end{align}
For fully degenerate free-particle masses $m_{1}=m_{2}$, this reduces to
\begin{align}
\label{series:2a}
\lim_{|\Delta m|^{2}\to0}\widetilde m_{1,2}^{s}(E,B)=m_{1}\mp\mathcal{C}_{s}\,.
\end{align}
Equation (\ref{series:2a}) indicates that for degenerate free-particle masses, the splitting $|\Delta m_\mathrm{EM}|\equiv\mathcal{C}_{s}$ between effective EM-masses arises purely from the electromagnetic interaction of the neutrinos. We return to this interesting insight in our final comments.

Because of the bounds in~\req{bound:1} on the neutrino magnetic moment, we can estimate the field strength required for an external magnetic field to generate an electromagnetic mass splitting of $|\Delta m_\mathrm{EM}|=10^{-3}$~eV which is a reasonable comparison to intrinsic splitting of the two similar massive neutrinos based on the experimental limits on neutrino masses~\cite{ParticleDataGroup:2022pth}. Using the upper limit for the neutrino magnetic moment of $\mu_{\nu}^\mathrm{eff}\sim10^{-10}\mu_{B}$ we obtain
\begin{align}
\label{estimate:1}
\left.\frac{\mathcal{C}_{s}}{\mu_{\nu}^\mathrm{eff}}\right\rvert_{\vec{E}=0}=\frac{10^{-3}\,\mathrm{eV}}{10^{-10}\mu_{B}}\approx1.7\times10^{11}\,\mathrm{T}\,.
\end{align}
This is near the upper bound of the magnetic field strength of magnetars~\cite{Kaspi:2017fwg} which are of the order $10^{11}$~Tesla. In this situation, the EM contribution to the mass splitting rivals the estimated inherent splitting~\cite{ParticleDataGroup:2022pth} of the two closer in mass neutrinos. Primordial magnetic fields~\cite{Grasso:2000wj} in the Early Universe may also present an environment for significant EM neutrino flavor mixing as both the external field strength and the density of neutrinos would be very large~\cite{Rafelski:2023emw}. The magnetic properties of neutrinos may also have contributed alongside the charged leptons in magnetization in the Early Universe~\cite{Steinmetz:2023nsc} prior to recombination. 

While the above estimate was done with astrophysical systems in mind, we note that strong electrical fields should also produce EM-mass splitting. Therefore environments near to high $Z$-nuclei also of interest~\cite{Bouchiat:1974kt,Bouchiat:1997mj,Safronova:2017xyt}, as weak interactions violate parity. Should neutrinos have abnormally large transition magnetic dipole moments, then they should exhibit mass splitting from the neutrino's electromagnetic dipole interaction which may compete with the intrinsic mass differences of the free-particles.

%%%%%%%%%%%%%%%%%%%%%%%%%%%%%%%%%%%%
\section{Conclusions}
\label{sec:conclusions}
%%%%%%%%%%%%%%%%%%%%%%%%%%%%%%%%%%%%
We have incorporated electromagnetic effects in the Majorana neutrino mixing matrix by introducing an anomalous transition magnetic dipole moment. We have described the formalism for three generations of neutrinos and explicitly explored the two generation case as a toy model. 

In the two generation case, we determined the effect of electric and magnetic fields on flavor rotation in~\req{zrot:2} by introducing an electromagnetic flavor unitary rotation $Z_{kj}^\mathrm{ext}$. We presented remixed mass eigenstates $\widetilde m(E,B)$ in~\req{poly:3} which are the propagating mass-states in a background electromagnetic field $F^{\alpha\beta}_\mathrm{ext}(x^{\mu})$. These EM-mass eigenstates were also further split by spin aligned and anti-aligned states relative to the external field momentum density. There is much left to do to explore further the nascent connection between the spin and the flavor via transition magnetic moments. 

Of particular interest is the case of nearly degenerate free-particle mass eigenstates where the EM effects are most manifest. For nearly degenerate masses compared to EM fields described by $\mathcal{C}_{s}(E,B)\gg|\Delta m|^{2}$ in~\req{spinsplit:1}, the mass splitting is dominated by the electromagnetic contribution. This effect could be most relevant in the strong magnetic field environments, such as around magnetars or primordial magnetic fields in the Early Universe, where it could be capable of competing with the mass splitting seen in the two closer neutrino mass states. We also emphasize that dense matter environments may also be relevant where the electrical field energy density is large. A natural extension of this work would be to include the in matter weak interaction remixing as an additional term in the effective Lagrangian~\req{massmom:2}.
 
The transition dipole moments are the origin of dynamical flavor mixing. While our focus was on Majorana neutrinos, Dirac-type fermions (neutrinos included) may also have non-zero transition  dipole moments. These could  remix flavor in the presence of strong external background fields. Here  quarks  are of special interest because they are not only electrically charged, but have color charge as well. This means quarks could in principle possess one, or both, EM and color-charge transition dipole moments, leading to dynamical effects in the CKM mixing matrix within hadrons as well as in quark-gluon plasma.

More speculatively, as transition dipoles act as a mechanism to generate mass by virtue of EM energy density $T_\mathrm{ext}^{00}$ as seen in~\req{poly:3}, an analogous consequence of our work could arise in the presence of a dark vector field in the Universe coupled to neutrinos, resulting in off-diagonal masses in flavor. Massless neutrinos could then obtain dynamical non-zero masses in the Universe by virtue of their interactions originating in dark transition moments.

%%%%%%%%%%%%%%%%%%%%%%%%%%%%%%%%%%%%
\section*{Acknowledgements}
\label{sec:acknowledgements}
%%%%%%%%%%%%%%%%%%%%%%%%%%%%%%%%%%%%
This article is dedicated to the memory of Harald Fritzsch. A version of this article with extended personal remarks appears in the book dedicated to the memory of Harald Fritzsch edited by Gerhard Buchalla, Dieter L\"ust and Zhi-Zhong Xing. With Harald's passing we lost a friend  and, equally importantly, a colleague whose quick mind, willingness to listen, and clarity of thought, helped in some of our research challenges. We had very much wished to hear Harald's opinion on this work.

%%%%%%%%%%%%%%%%%%%%%%%%%%%%%%%%%%%%
\bibliographystyle{unsrturl}
\bibliography{neutrino-transition-moments-refs}
%%%%%%%%%%%%%%%%%%%%%%%%%%%%%%%%%%%%
\end{document}